\documentclass[11pt,twoside]{article}  
\usepackage{asp2006}
\usepackage{adassconf}
\usepackage{graphicx}

\newcommand{\slang}{S-Lang}
\newcommand{\omp}{OpenMP}
\newcommand{\isis}{ISIS}
\newcommand{\slirp}{SLIRP}

\newcommand{\lint}{{\rm Linux2}}
\newcommand{\solf}{{\rm Solaris4}}

\begin{document}   

\paperID{P10.5}

\title{Getting More From Your Multicore: Exploiting OpenMP for Astronomy}

%
%
%
%
%

\markboth{M.S. Noble}{Getting More From Your Multicore}

%
%
%
%

\author{Michael S. Noble}
\affil{Kavli Institute for Astrophysics,Massachusetts Institute of Technology}

%

\contact{Mike Noble}
\email{mnoble@space.mit.edu}

%
%
%

\paindex{Noble, M.S.}

%



\begin{abstract}          
Motivated by the emergence of multicore architectures, and the reality
that parallelism is rarely used for analysis in observational astronomy,
we demonstrate how general users may employ tightly-coupled
multiprocessors in scriptable research calculations while requiring no
special knowledge of parallel programming.
Our method rests on the observation that much of the appeal of high-level
vectorized languages like IDL or MatLab stems from relatively simple
internal loops over regular array structures, and that these loops are
highly amenable to automatic parallelization with OpenMP.
We discuss how ISIS, an open-source astrophysical analysis system
embedding the \slang\ numerical language, was easily adapted to exploit
this pattern.  Drawing from a common astrophysical problem, model
fitting, we present beneficial speedups for several machine and compiler
configurations.  These results complement our previous efforts with PVM,
and together lead us to believe that ISIS is the only general purpose
spectroscopy system in which such a range of parallelism -- from single
processors on multiple machines to multiple processors on single
machines -- has been demonstrated.
\end{abstract}

%
%

\section{Problem: Underpowered Analysis}
As noted in Noble et al (2006), parallel computation is barely
used in astronomical analysis. For example, models in XSPEC (Arnaud 1996),
the de facto standard X-ray spectral analysis tool, still run serially
on my dual-CPU desktop. In this situation scientists tend to either
turn away from models which are expensive to compute or just accept
that they will run slowly.  Analysis systems which do not embrace
parallelism can process at most the workload of only 1 CPU, resulting
in a dramatic $1/n$ underutilization of resources as more CPU cores
are added.
At the same time, however, astronomers are well versed in scripting,
particularly with very high-level, array-oriented numerical packages
like IDL, PDL, and \slang, to name a few.
They combine easy manipulation of mathematical structures of arbitrary
dimension with most of the performance of compiled code, with the latter
due largely to moving array traversals from the interpreted layer into
lower-level code like this C fragment
{\small
\begin{verbatim}
                  case SLANG_TIMES:
                       for (n = 0; n $<$ na; n++)
                           c[n] = a[n] * b[n];
\end{verbatim}
}
\noindent
which provides vectorized multiplication in \slang.
This suggests that much of the strength and appeal of numerical scripting
stems from relatively simple loops over regular structures.  Having
such loops in lower-level
compiled codes also makes them ripe for parallelization with \omp\ on
shared memory multiprocessors.  Proponents contend that conceptual
simplicity makes \omp\ more approachable
than other parallel programming models, e.g. message-passing in MPI or PVM,
and emphasize the added benefit of allowing single bodies of code to be used
for both serial and parallel execution.  For instance, preceding the above
loop with \verb+#pragma omp parallel for+
parallelizes the \slang\ multiplication operator; the pragma is simply
ignored by a non-conforming compiler, resulting in a sequential program.

\section{Parallelizing ISIS by way of OpenMP in SLIRP and S-Lang}
\isis\ (Houck, 2002) was conceived to support analysis of high-resolution
Chandra X-Ray
gratings spectra, then quickly grew into a general-purpose analysis system;
it is essentially a superset of XSPEC, combining all of its models and more
with the \slang\ scripting language, whose mathematical capabilities rival
commercial
packages such as MatLab or IDL.  One of several distinguishing features
possessed by \isis\ is its ability to bring the aggregate power of
workstation clusters, through a fault-tolerant PVM interface, to bear on
general problems in spectroscopic analysis (Noble et al 2006).  This paper
complements that work by discussing two ways in which \omp\ has been used
to enable shared-memory parallelism in \isis.
Like our PVM work, we believe this is another first for general-purpose
X-ray spectroscopy, and is of added significance in that adapting \isis\
for parallelism -- both distributed and shared-memory -- no modifications
to its architecture or internal codebase were required.

\begin{figure*}
 \centering
 \includegraphics[angle=-90,scale=0.22]{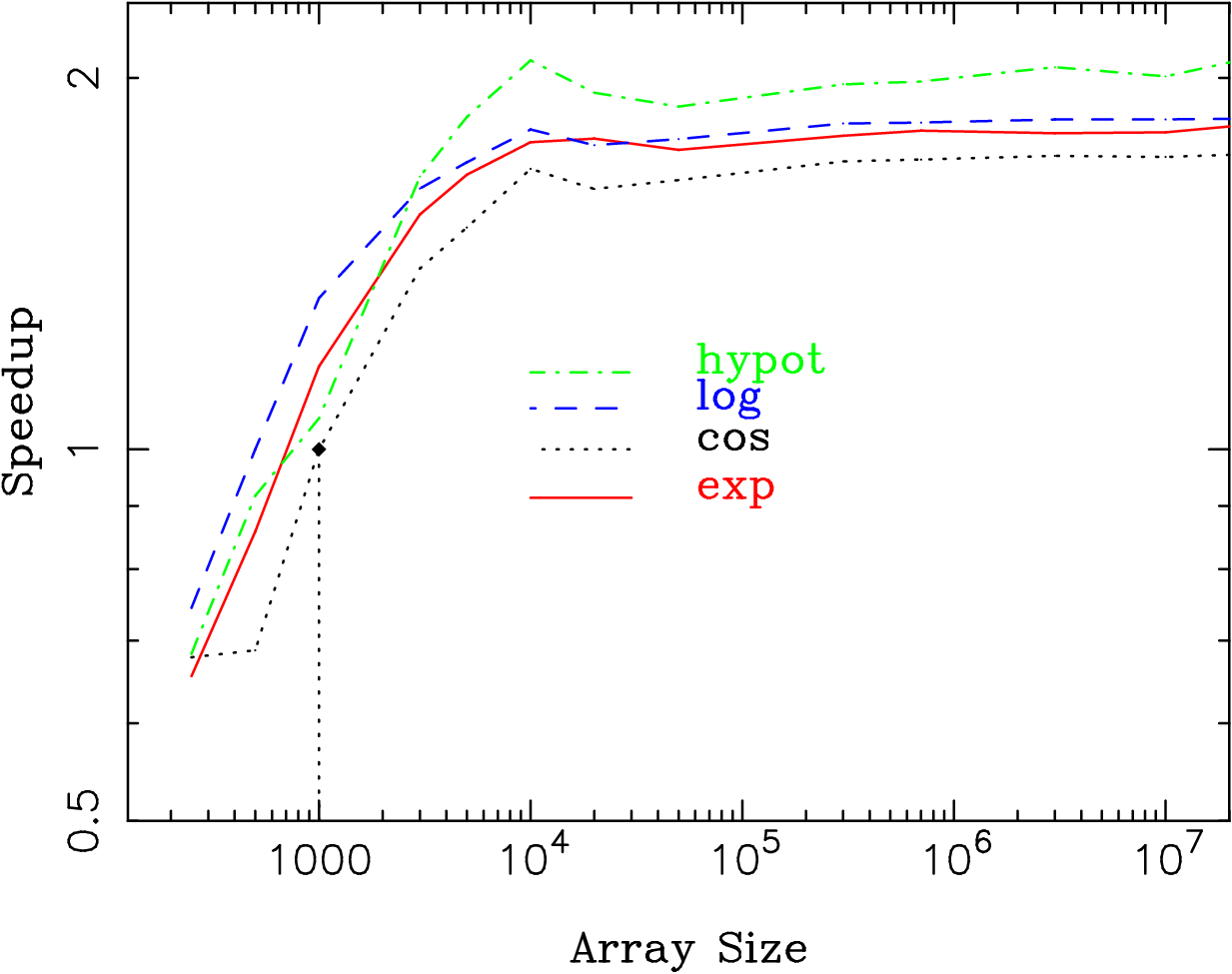}
 \hspace*{6mm}
 \includegraphics[angle=-90,scale=0.22]{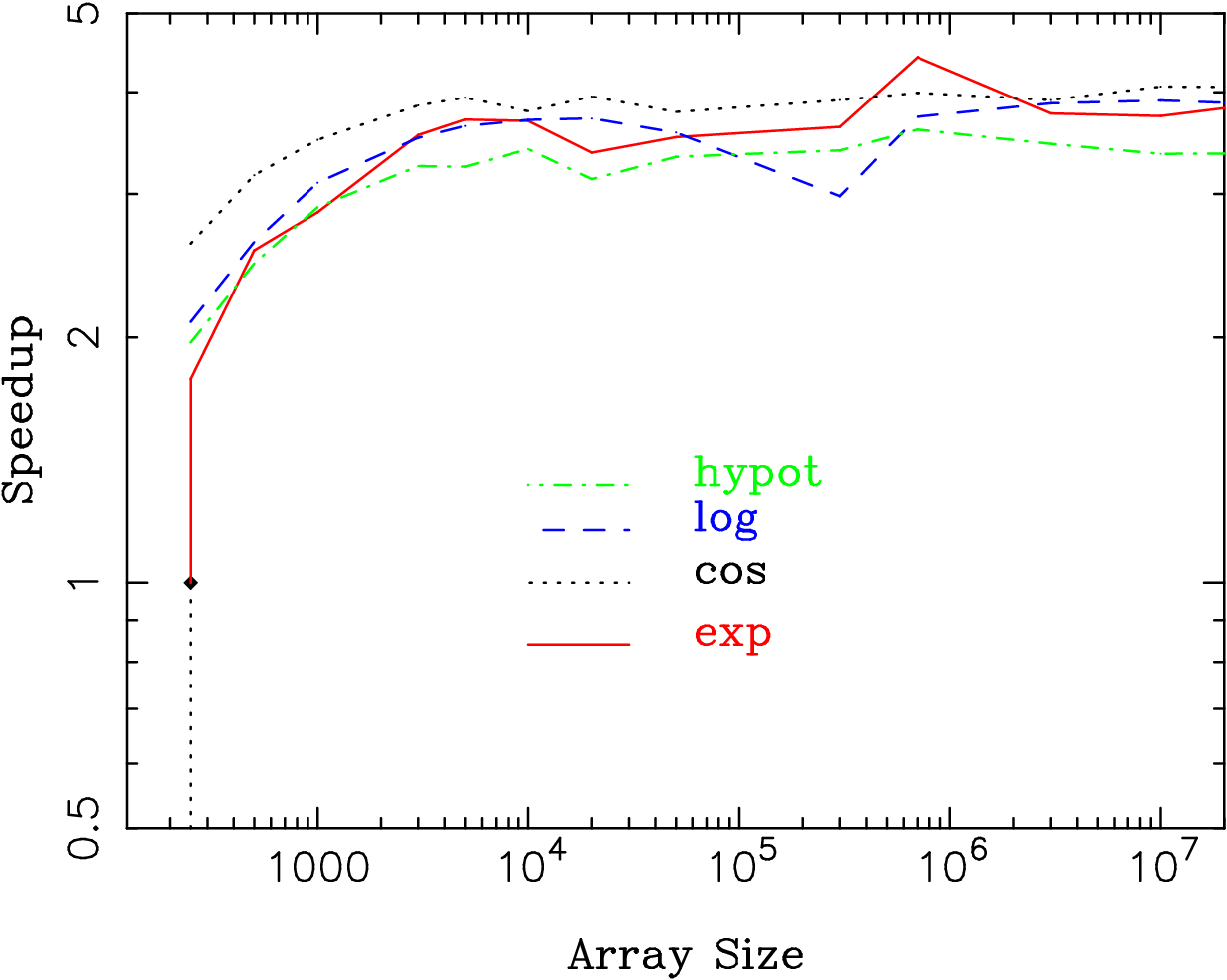}
 \caption{Speedups after replacing selected \slang\ math intrinsics with
 parallelized versions generated by \slirp.
 Left: 2 CPUs on Linux (\lint).  Right: 4 CPUs on Solaris (\solf).
 The dotted vertical lines mark the inflection points where parallel
 performance begins to overtake serial.
 }
 \label{P10.5-fig-1}
\end{figure*}

\begin{figure*}[t]
 \centering
 \includegraphics[angle=-90,scale=0.22]{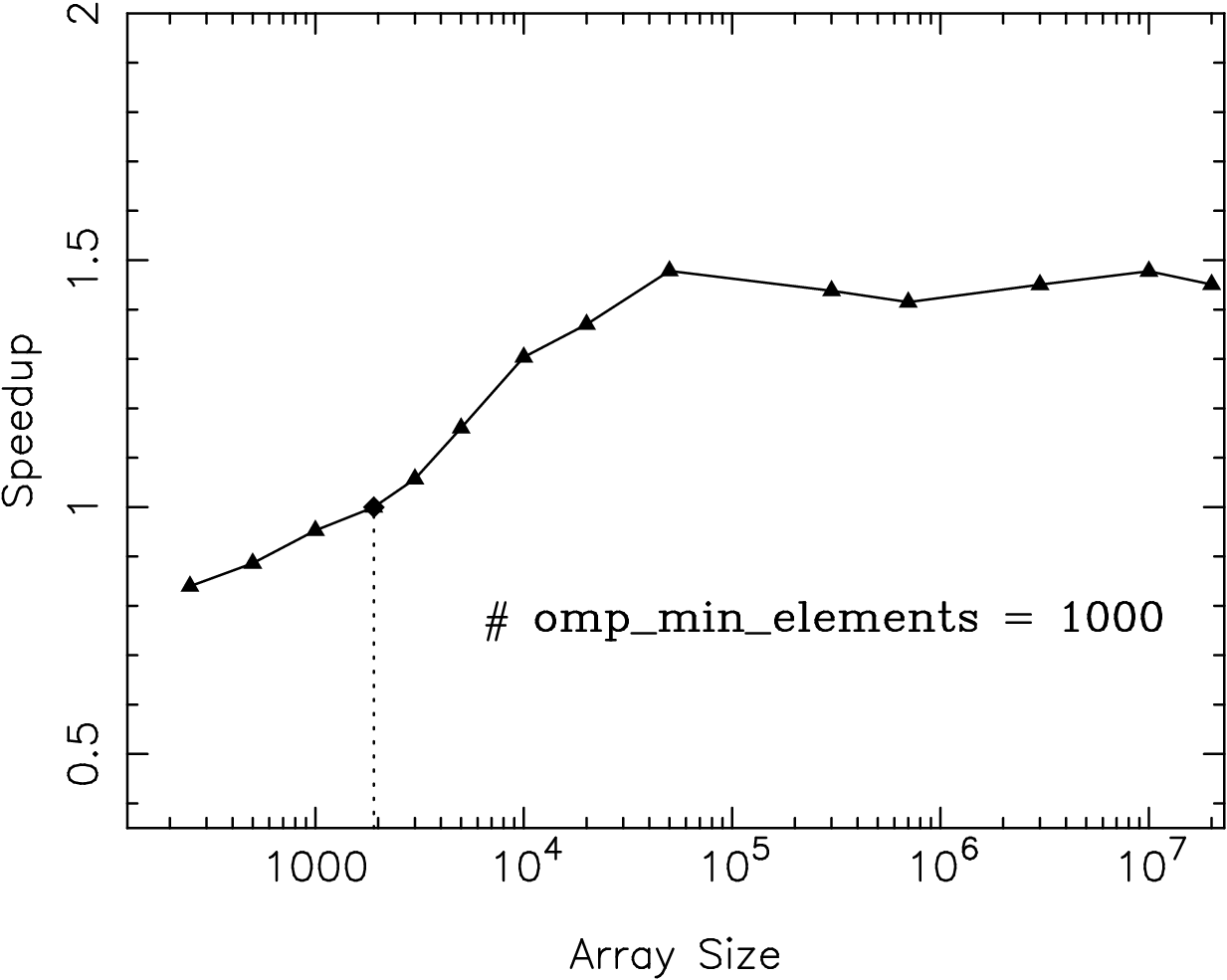}
 \hspace*{6mm}
 \includegraphics[angle=-90,scale=0.22]{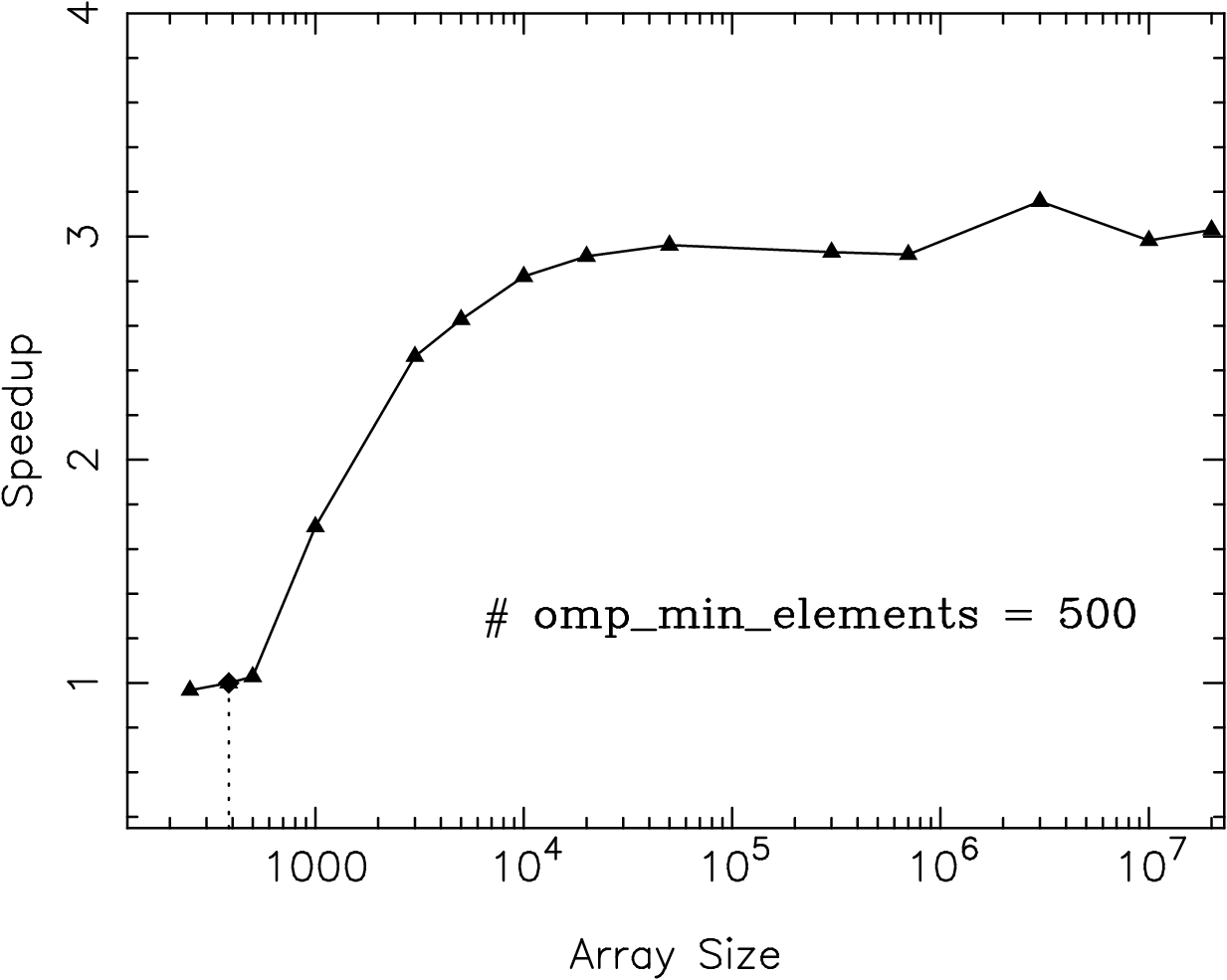}
 \caption{Aggregate speedup of the Weibull fit function due to the
 parallelized operators and functions discussed above.  Left: \lint,
 with inflection point at 1907 elements.  Right: \solf, with
 inflection point at 384 elements.
 }
 \label{P10.5-fig-2}
\end{figure*}

The first manner in which \isis\ was endowed with multicore capability
involved loading a module of parallelized wrappers for C functions
such as {\tt exp} and {\tt hypot} (Fig. \ref{P10.5-fig-1}).  These bindings
were created by \slirp, which is distinguished by its ability to auto-generate
vectorized wrappers such as
{\small
\begin{verbatim}
            static void sl_atof (void)
            {
               ...
               for (_viter=0; _viter < vs.num_iters; _viter++)
                   retval[_viter] = atof((char*)*arg1);
               ...
            }
\end{verbatim}
}
\noindent
that allow functions which ordinarily accept only scalar inputs to also
be used with array semantics.  Adapting \slirp\ to enable these vectorized
wrappers to run in parallel was as simple as having it prefix the
vectorization loop with an \omp\ { \tt \#pragma} as detailed above.
This enables \slang\ intrinsics, all of which execute serially, to be
replaced with parallel versions of the same name, transparently parallelizing
the replaced functions.
Our second multicore tactic involved minimizing the effects of Amdahl's law
by parallelizing a number of \slang\ operators (and part of the {\tt where()}
function), through the addition of
{\small
\begin{verbatim} 
           #pragma omp parallel for if (size > omp_min_elements)
\end{verbatim}
}
\noindent
to the corresponding loops within the \slang\ interpreter source
The {\tt if} clause in these
directives was used to tune performance for small arrays, where
the cost of threads outweighs the serial execution
time.
\noindent
The speedup plots in Figs. \ref{P10.5-fig-1} and \ref{P10.5-fig-2}
demonstrate significant performance gains from the use of shared-memory
parallelism in \isis.  Our numbers represent measurements
of prerelease GCC 4.2 -O2 builds on 2 CPUs running Debian Linux (Linux2)
and Sun Studio 9 -xO3 builds on 4 CPUs running Solaris (Solaris4), with
position independent compilation.
Performance of the parallelized functions approaches the theoretical
maximum of linear speedup as array sizes increase (Fig. \ref{P10.5-fig-1}),
and the
inflection points in the size of the arrays needed for nominal speedup
from multithreading (represented by the dotted vertical lines) are
relatively small, ca. 1000 elements on \lint\ and 250 elements on
\solf.\footnote{More performance numbers and discussion, as well as the
code for the Weibull function discussed herein, are given in the arXiv
e-print at http://arxiv.org/abs/0706.4048}
\section{Case Study: Weibull Model in ISIS Spectroscopy}
While \isis\ supports custom user models in Fortran and C/C++, it can be
faster to code them directly in \slang\ and avoid compilation steps during
experimental tuning. In this section we discuss how one such function, a
4-parameter Weibull model coded for serial execution and taken directly
from an active research project at MIT, parallelized using the techniques
detailed above.  Fig. \ref{P10.5-fig-2} shows realized speedups converging
on ca. 150\% for \lint\ and 300\% for \solf.
These are sizable performance increases, and especially significant
in that end-users need to do nothing -- in terms of learning parallelism
or recoding sequential algorithms -- to obtain them; the same top-level
model script can be used for both parallel and serial execution.
Furthermore, recall that these models are used in the context of an
iterative fitting process.  Fits do not converge after just one iteration,
and generating accurate confidence intervals -- an absolute necessity for
credible modeling -- can require that tens or hundreds of thousands of
fits be performed at each point on a parameter space grid.  In such cases
the speedups given here accumulate to significant differences in the
overall runtime of an analysis sequence.

\section{Conclusion: Transparently Parallel Scripting}
We are witnessing the arrival of serious multiprocessing capability on
the desktop: multicore chip designs are making it possible for general
users to access many processors. At the granularity of the operating
system it will be
relatively easy to make use of these extra cores, say by assigning whole
programs to separate CPUs.  As noted with increasing frequency of late,
though, it is not as straightforward to exploit this concurrency {\em within}
individual desktop applications.  In this paper we demonstrated how we
have helped our research colleagues prepare for this eventuality.
We have enhanced the vectorization capabilities of \slirp, a module
generator for the \slang\ numerical scripting language, so that wrappers
may be annotated for automatic parallelization with \omp.  This lets
\slang\ intrinsic functions be replaced with parallelized
versions, at runtime, without modifying a single line of internal \slang\
source.  We have shown how \slang\ operators may also be parallelized with
relative ease, by identifying key loops within the interpreter source,
tagging them with \omp\ directives and recompiling.  These simple adaptations,
which did not require any changes to the \isis\ architecture or codebase,
have yielded beneficial speedups for computations actively used in
astrophysical research, and allow the same numerical scripts to be used
for both serial and parallel execution --
minimizing two traditional barriers to the use of parallelism by
non-specialists: learning how to program for concurrency and recasting
sequential algorithms in parallel form.  By transparently using
\omp\ to effect greater multiprocessor utilization we gain the freedom
to explore on the desktop more challenging problems that other researchers
might avoid for their prohibitive cost of computation.
The \omp\ support now available in GCC makes the techniques espoused here
a viable option for many open source numerical packages, opening the door
to wider adoption of parallel computing by general practitioners.

\acknowledgments
{\footnotesize
This work was supported by NASA through the Hydra AISRP grant NNG06GE58G,
and by contract SV-1-61010 from the Smithsonian Institution.}

\end{document}